\documentclass{article}
\usepackage{amsfonts,amssymb, amsmath}
\usepackage[english]{babel}

\def\nn{\nonumber }
\def\bq{ \begin{equation} }
\def\eq{ \end{equation} }
\def\ben{ \begin{eqnarray} }
\def\en{ \end{eqnarray} }
\def\e{{\rm e}}
\def\ii{{\rm i}}

\newtheorem{prop}{Proposition}

\newtheorem{defi}{Definition}

\newtheorem{re}{Remark}
\newenvironment{rem}{\begin{re} \rm }{\end{re}}

\begin{document}


\title{On natural Poisson bivectors on the sphere}
\author{ A V Tsiganov \\
\it\small
St.Petersburg State University, St.Petersburg, Russia\\
\it\small e--mail: andrey.tsiganov@gmail.com}

\date{}
\maketitle

\begin{abstract}
 We discuss the concept of natural Poisson bivectors, which allows us to consider the overwhelming majority of known integrable systems on the sphere in framework of bi-Hamiltonian geometry.
\end{abstract}

\vskip0.1truecm
\section{Introduction}
\setcounter{equation}{0}
The Hamilton-Jacobi theory seems to be one of the most powerful methods of investigation the dynamics of mechanical (holonomic and nonholonomic) and control systems. Besides its fundamental aspects such as its relation to the action integral and generating functions of symplectic maps, the theory is known to be very useful in integrating the Hamilton equations using the variables separation technique. The milestones of this technique include the works of St\"{a}ckel, Levi-Civita, Eisenhart, Woodhouse, Kalnins, Miller, Benenti and others. The majority of results was obtained for a very special class of integrable systems, important from the physical point of view, namely for the systems with quadratic in momenta integrals of motion. The Kowalevski, Chaplygin and Goryachev results on separation of variables for the systems with higher order integrals of motion missed out of this scheme.

Bi-Hamiltonian structures can be seen as a dual formulation of integrability and separability, in the sense that they substitute a hierarchy of compatible Poisson structures to the hierarchy of functions in involution, which may be treated either as integrals of motion or as variables of separation for some dynamical system. The Eisenhart-Benenti theory was embedded into the bi-Hamiltonian set-up using the lifting of the conformal Killing tensor that lies at the heart of Benenti's construction \cite{sar00,imm00}. The concept of natural Poisson bivectors allows us to generalize this construction and to study systems with quadratic and higher order integrals of motion in framework of a single theory \cite{ts10d}.

The aim of this note is to bring together all the known examples of natural Poisson bivectors on the sphere, because a good example is the best sermon. Some of these Poisson bivectors have been obtained and presented earlier in different coordinate systems and notations. Here we propose the unified description of this known and few new bivectors  using so-called geodesic $\Pi$ and potential $\Lambda$ matrices \cite{ts10d}. In some sense we propose new form for the old content and believe that this unification is a first step to the geometric analysis of various natural systems on the sphere, which reveals what they have in common and indicates the most suitable strategy to obtain and to analyze their solutions.

The corresponding integrable natural systems on two-dimensional unit
sphere $\mathbb S^2$ are related to rigid body dynamics. In order to describe these systems we will use the angular momentum vector $J=(J_1,J_2,J_3)$ and the Poisson vector $x=(x_1,x_2,x_3)$ in a moving frame of coordinates attached to the principal axes of inertia \cite{bm05}. The Poisson brackets between these variables
\begin{equation}\label{e3}
\,\qquad \bigl\{J_i\,,J_j\,\bigr\}=\varepsilon_{ijk}J_k\,, \qquad
\bigl\{J_i\,,x_j\,\bigr\}=\varepsilon_{ijk}x_k \,, \qquad
\bigl\{x_i\,,x_j\,\bigr\}=0\,,
\end{equation}
may be associated to the Lie-Poisson algebra of the three-dimensional Euclidean algebra $e(3)$ with two Casimir elements
\bq \label{caz-e3}
 C_1=|x|^2\equiv\sum_{k=1}^3 x_k^2, \qquad C_2= \langle x,J \rangle\equiv\sum_{k=1}^3 x_kJ_k .
\eq
Below we always put $C_2=0$.

As usual all the results are presented up to the linear canonical transformations, which consist of rotations
\[
 x\to \alpha\, {U}\, x\,,\qquad J\to {U}\, J\,,
\]
where $\alpha$ is an arbitrary parameter and $U$ is an orthogonal constant matrix,
and shifts
\[
x\to x \,,\qquad J\to J+ {S}\, x\,,\label{shiftE3}
\]
where ${ S}$ is an arbitrary $3\!\times\!3$ skew-symmetric constant matrix \cite{bm05,kst03}.

 If the square integral of motion $C_2=(x,J)$ is equal to zero, rigid body dynamics may be restricted on the unit sphere $\mathbb S^2$ and we can use standard spherical coordinate system on it's cotangent bundle $T^*{\mathbb S}^2$
\bq\label{sph-coord}
\begin{array}{lll}
x_1 =\sin\phi\sin\theta,\qquad& x_2 = \cos\phi\sin\theta,\qquad & x_3 =\cos\theta\,,\\
\\
J_1 =\dfrac{\sin\phi\cos\theta}{\sin\theta}\,p_\phi-\cos\phi\,p_\theta\,,\qquad&
J_2 =\dfrac{\cos\phi\cos\theta}{\sin\theta}\,p_\phi+\sin\phi\,p_\theta\,,
\qquad& J_3 = -p_\phi\,.
\end{array}
\eq
We use these variables in order to determine and classify the natural Poisson bivectors on $T^*\mathbb S^2$ up to the point canonical transformations.

As far as the organization of this paper is concerned, in Section 2 we briefly introduce the notions of bi-Hamiltonian geometry relevant for subsequent sections. In particular, we discuss the concept of natural Poisson bivectors on cotangent bundles to Riemannian manifolds, which allows us to generalize classical Eisenhart-Benenti theory. In Section 3 we discuss the bi-Hamiltonian classification of bi-integrable systems on the sphere. Section 4 is devoted to the separable natural systems coming from auxiliary bi-Hamiltonian systems.

\section{Some issues in the geometry of bi-Hamiltonian manifolds}
\setcounter{equation}{0}

A bi-Hamiltonian manifold $M$ is a smooth manifold endowed
with a pair of compatible Poisson bivectors $P$ and $P'$ such that
\bq\label{m-eq1}
[P,P']=0,\qquad [P',P']=0,\qquad
\eq
where $[.,.]$ is the Schouten bracket. This means that every linear combination of $P$ and $P'$ is still a Poisson bivector.

If $P$ is invertible Poisson bivector on $M$, one can introduce the so-called Nijenhuis operator (or hereditary, or recursion)
\bq\label{rec-op}
N=P' P^{-1}\,.
\eq
If $N$ has, at every point, the maximal number
of different functionally independent eigenvalues $u_1,\ldots,u_n$, then $M$ is said to be a regular bi-Hamiltonian manifold.

\subsection{Bi-integrable systems }
Let us consider a family of bi-integrable systems for which
 there are functionally independent integrals of motion $H_1,\ldots,H_n$ in the bi-involution
\bq\label{bi-inv}
\{H_i,H_j\}=\{H_i,H_j\}'=0\,,\qquad i,j=1,\ldots,n,
\eq
with respect to a pair of compatible Poisson brackets $\{.,.\}$ and $\{.,.\}'$ defined by $P$ and $P'$.
 There are three known distinct constructions of bi-integrable systems, see \cite{ts10d} .

Firstly, if $M$ is a regular bi-Hamiltonian manifold endowed with invertible Poisson bivector $P$, then we can construct
 recursion operator $N$ (\ref{rec-op}) and, as usual, functions
\begin{equation}\label{aux-int}
\mathcal H_k=\frac1{2k}\,\mathrm{tr}\,N^k
\end{equation}
form a bi-Hamiltonian hierarchy on $M$, i.e. the Lenard relations hold
\[P' d{\mathcal H}_k=P d{\mathcal H}_{k+1}\,,\qquad\mbox{for all}\qquad k\ge 1\,.
 \]
Using these relations we can get all the integrals of motion starting with the Hamilton function $H_1$.
\begin{rem}
 The natural obstacle for existence of the bi-Hamiltonian systems is discussed in \cite{br}. Fortunately, we can use these rare bi-Hamiltonian systems (natural or non-natural) as  \textit{auxiliary} systems for the construction of an infinite family of non bi-Hamiltonian separable systems.
\end{rem}

Namely, second special but more fundamental construction of integrable systems was originally formulated by Jacobi when he invented elliptic coordinates and successfully applied them to solve several important mechanical problems:
\textit{"The main difficulty in integrating a given differential equation lies in introducing convenient variables, which there is no rule for finding. Therefore, we must travel the reverse path and after finding some notable substitution, look for problems to which it can be successfully applied".}

In framework of the Jacobi method we consider $\mathcal H_i$ (\ref{aux-int}) as constants of motion for an \textit{auxiliary} bi-Hamiltonian system on the regular bi-Hamiltonian manifold $M$ and treat functionally independent eigenvalues $u_j$ of $N$
\bq\label{dn-var}
B(\lambda)=\Bigl(\,\det(N-\lambda\mathrm I)\,\Bigr)^{1/2}=(\lambda-u_1)(\lambda-u_2)\cdots
(\lambda-u_n)\,,
\eq
 as "convenient variables" for an ifinite family of \textit{separable} bi-integrable systems associated with various separated relations
 \begin{equation}
\label{seprel}
\Phi_i(u_i,p_{u_i},H_1,\dots,H_n)=0\ ,\quad i=1,\dots,n\ ,
\qquad\mbox{with }\det\left[\frac{\partial \Phi_i}{\partial H_j}\right]
\not=0\> .
\end{equation}
Here $u=(u_1,\ldots,u_n)$ and $p_u=(p_{u_1},\ldots,p_{u_n})$ are canonical variables of separation
\bq\label{poi-br12}
\{u_i,p_{u_j}\}=\delta_{ij}\,\qquad\mbox{and}\qquad \{u_i,p_{u_j}\}'=\delta_{ij}\,u_i\,.
\eq
The Poisson brackets (\ref{poi-br12}) entail that solutions $H_1,\ldots,H_n$ of the separated relations (\ref{seprel}) are functionally independent integrals of motion in the bi-involution (\ref{bi-inv}), see \cite{ts07a}. Of course, this construction will be justified only if we are capable to obtain separable Hamilton functions $H=H_i$, which have natural form in initial $(p,q)$ variables (\ref{nat-h}).

The third construction of integrals of motion in bi-involution on irregular bi-Hamiltonian manifolds is discussed in \cite{mpt,ts10d}. In this case polynomial integrals of motion $H_2,\ldots,H_n$ are solutions of the following equations for the given Hamiltonian $H_1$
\[
P'dH_1=\varkappa_k P\,d\ln H_k,\qquad k> 1\,, \quad \varkappa_k\in\mathbb R\,,
\]
which replace the usual Lenard relations (\ref{aux-int}). If this equations have many different functionally independent solutions labeled by different $\varkappa_k$, then we obtain so-called superintegrable systems \cite{mpt,ts10d}.

\subsection{Bi-Hamiltonian structures on cotangent bundles}
According to \cite{Turiel} a torsionless (1,1) tensor field $L$ on a smooth manifold $Q$ gives rise to a (second) Poisson structure on the cotangent space $M=T^*Q$, compatible with the canonical one.

Let $\theta$ be the Liouville $1$-form on $T^*Q$ and $\omega= d\theta$ the standard symplectic
$2$-form on $T^\ast Q$, whose associated Poisson bivector will be denoted with $P$.
If we choose some local coordinates $q=(q_1,\dots,q_n)$ on $Q$ and the corresponding symplectic coordinates $(q,p)=(q_1,\dots,q_n,p_1,\dots,p_n)$ on
$T^\ast Q$ then we get the following local expressions
 \bq\label{poi-0}
 \theta=p_1dq_1+\ldots p_ndq_n\,,\qquad\mbox{and}\qquad P=\left(
 \begin{array}{cc}
 0 & \mathrm{I} \\
 -\mathrm{I} & 0
 \end{array}
\right)\,.\eq
Using a torsionless tensor field $L$ one can deform $\theta$ to a $1$-form $\theta'$ and $P$ to bivector $P'$:
\bq \label{p2-ben}
\theta' =\sum_{i,j=1}^n L_{ij} p_i dq_j,\qquad\mbox{and}\qquad
 P'= \left(
 \begin{array}{cc}
 0 & L_{ij} \\
 \\
 -L_{ij}\qquad&\displaystyle \sum_{k=1}^n\left(\dfrac{\partial L_{ki}}{\partial q_j}-\dfrac{\partial L_{kj}}{\partial q_i}\right)p_k
 \end{array}
 \right)\,.
 \eq
The vanishing of $L$ torsion entails that $P'$ (\ref{p2-ben}) is a Poisson bivector compatible with $P$.

Let us consider natural integrable by Liouville system on $Q$.
\begin{defi} The natural  Hamilton function
\bq \label{nat-h}
H_1=T+V=\sum_{i,j=1}^n \mathrm g_{ij}\,p_i\,p_j+V(q_1,\ldots, q_n)\,
\eq
is the sum of the geodesic Hamiltonian $T$ defined by metric tensor $\mathrm g(q_1,\ldots,q_n)$ and potential energy $V(q_1,\ldots,q_n)$ on $Q$.
\end{defi}
 If the corresponding Hamilton-Jacobi equation is separable in orthogonal coordinate system $(u,p_u)$ on configurational space  $Q$,
 then in framework of the Eisenhart-Benenti theory the second Poisson bivector $P'$ (\ref{p2-ben}) is defined by a conformal Killing tensor $L$  of gradient type on $Q$ with pointwise simple eigenvalues associated with the metric $\mathrm g(q_1,\ldots,q_n)$, see \cite{ben97, ben05,bm03,sar00,imm00}.

According to Kowalevski \cite{kow89} and Chaplygin \cite{chap04}, separation of variables for integrable systems with higher order integrals of motion involves generic canonical transformation of the whole phase space. Definition  (\ref{nat-h})  of the natural Hamiltonian and metric tensor  $\mathrm g(q_1,\ldots,q_n)$ is non-invariant with respect to arbitrary canonical transformations of coordinates on $T^*Q$
\[
q_i\to q'_i=f_i(p,q)\,,\qquad p_i\to p'_i=g_i(p,q)\,.
\]
In the situation, when habitual objects (geodesic, metric, potential) lose their geometric sense and  remaining invariant equation (\ref{m-eq1}) has apriority infinite many solutions,  notion of the natural Poisson bivectors on $T^*Q$ became de-facto  very useful practical tool for the calculation of  variables of separation \cite{gts10,mpt,ts10a,ts10k,ts10d,ts09v}.
\begin{defi}
The natural Poisson bivector $P'$ on $T^*Q$ is a sum of the geodesic Poisson bivector $P'_T$ compatible with $P$
\bq\label{w-eq}
[P,P'_T]=[P'_T,P'_T]=0\,,
\eq
 and the potential part defined by a torsionless (1,1) tensor field $\Lambda(q_1,\ldots, q_n)$ on $Q$
\bq\label{n-p}
P'= P'_T+\left(
 \begin{array}{cc}
 0 & \Lambda_{ij} \\
 \\
 -\Lambda_{ji}\quad &\displaystyle \sum_{k=1}^n\left(\dfrac{\partial \Lambda_{ki}}{\partial q_j}-\dfrac{\partial \Lambda_{kj}}{\partial q_i}\right)p_k
 \end{array}
 \right)\,.
\eq
 \end{defi}
 In fact, here we simple assume that bi-integrability of the geodesic motion is a necessary condition for bi-integrability in generic case at $V\neq 0$.

Throughout this paper geodesic bivector $P'_T$ is defined by $n\times n $ matrix $\Pi(q_1,\ldots, q_n,p_1,\ldots,p_n)$ and functions $\mathrm{x,y}$ and $\mathrm z$ on $T^*Q$
\bq\label{p2-sph2}
 P'_T=
 \left(
 \begin{array}{cc}
 \displaystyle \sum_{k=1}^n\left( \mathrm{x}_{jk}(q)\dfrac{\partial \Pi_{jk}}{\partial p_i}-\mathrm{y}_{ik}(q)\dfrac{\partial \Pi_{ik}}{\partial p_j}\right) & \Pi_{ij} \\
 \\
 -\Pi_{ji}\quad&\displaystyle \sum_{k=1}^n\left(\dfrac{\partial \Pi_{ki}}{\partial q_j}-\dfrac{\partial \Pi_{kj}}{\partial q_i}\right)\,\mathrm z_{k}(p)\\
 \end{array}
 \right)
\,
 \eq
 up to the point transformations. In this case the corresponding Poisson bracket $\{.,.\}'$ looks like
\ben
&&\{q_i,p_j\}'=\Pi_{ij}+\Lambda_{ij}\,,\quad
\{q_i,q_j\}' = \displaystyle \sum_{k=1}^n\left( \mathrm{x}_{jk}(q)\dfrac{\partial \Pi_{jk}}{\partial p_i}-\mathrm{y}_{ik}(q)\dfrac{\partial \Pi_{ik}}{\partial p_j}\right) \,,\nn\\
&&\{p_i,p_j\}'=\sum_{k=1}^n\left(\dfrac{\partial \Lambda_{ki}}{\partial q_j}-\dfrac{\partial \Lambda_{kj}}{\partial q_i}\right)p_k+\sum_{k=1}^n\left(\dfrac{\partial \Pi_{ki}}{\partial q_j}-\dfrac{\partial \Pi_{kj}}{\partial q_i}\right)\,\mathrm z_{k}(p\,)\,.\nn
\en
In fact, functions $\mathrm{x,y}$ and $\mathrm z$ are completely determined by the matrix $\Pi$ via compatibility conditions (\ref{w-eq}).

 We can add various integrable potentials $V$ to the given geodesic Hamiltonian $T$ in order to get integrable natural Hamiltonians (\ref{nat-h}). In similar manner we can add different compatible potential matrices $\Lambda$ to the given geodesic matrix $\Pi$ in order to get natural Poisson bivectors $P'$ (\ref{n-p}) compatible with the canonical bivector $P$.
\begin{rem}
We have to underline that this definition of natural Poisson bivectors is the useful anzats rather than rigorous mathematical definition. It is an obvious sequence of non-invariant definition of the natural Hamiltonian  with respect to transformations of the whole phase space. We hope that further inquiry of geometric relations between $n\times n$ metric matrix $\mathrm g$, potential matrix $\Lambda$ and geodesic matrix $\Pi$ on $T^*Q$ allows us to get more invariant and rigorous definition of these objects.
\end{rem}

\begin{rem}
In term of variables of separation $\Pi=0$ and $\Lambda=\mbox{diag}(u_1,\ldots, u_n)$, so  we have usual invariant construction of Turiel \cite{Turiel}. The main problem is how to rewrite this invariant theory in term of initial physical variables.
\end{rem}

\begin{rem}
We suppose that (\ref{p2-sph2}) is a special form of $P'_T$. Other form of $P'_T$ on the generic symplectic leaves of $e^*(3)$ for the Steklov-Lyapunov system at $C_2\neq 0$ will be presented in the forthcoming publication.
\end{rem}

\section{Special natural Poisson bivectors on the sphere}
\setcounter{equation}{0}
The standard Laplace method for the direct search of integrable systems may be applied to the search of the
natural bivectors $P'$ too.

Firstly we stint ourselves by a family of natural Poisson bivectors (\ref{n-p}) with geodesic part (\ref{p2-sph2}).
Then, it is easy to see that the geodesic Hamiltonian
\[T= \sum_{i,j=1}^n \mathrm g_{ij}(q)\,p_i\,p_j\,\]
 on the cotangent bundle $T^*Q$ is the second order homogeneous polynomial in momenta, so we assume that entries of $\Pi$ are the similar homogeneous polynomials
\bq\label{pi2-sph}
\Pi_{ij}=\sum_{k,m=1}^n c_{ij}^{km}(q) p_kp_m
\eq
up to canonical transformations $p_k\to p_k+f_k(q_k)$. On two-dimensional unit sphere $Q=\mathbb S^2$ we use spherical coordinates (\ref{sph-coord}) such that
 \bq
 q=(q_{1},q_{2})=(\phi,\theta)\qquad\mbox{and}\qquad p=(p_1,p_2)=(p_\phi,p_\theta)\,.\label{p12-coord}
 \eq
 At the third step we introduce a family of partial solutions for which all the entries of $P'_T$ (\ref{p2-sph2}) are independent on variable $\phi$, i.e. at
\bq\label{assum2}
c_{ij}^{km}(q)=c_{ij}^{km}(\theta),\qquad \mathrm x_{jk}(\phi,\theta)=\mathrm x_{jk}(\theta)\,,\qquad
\mathrm y_{ik}(\phi,\theta)=\mathrm y_{ik}(\theta)\,.
\eq
It looks like reasonable assumption because the geodesic Hamiltonian $T$
\ben T&=&a_1J_1^2+a_2J_2^2+a_3J_3^2=\Bigl(a_3-\cot^2\theta(a_1\sin^2\phi+a_2\cos^2\phi)\Bigr)p_\phi^2\,,\nn\\
\label{n-hams}\\
&-&\sin2\phi\cot\theta(a_1-a_2)p_\theta\,p_\phi+(a_1\cos^2\phi+a_2\sin^2\phi)p_\theta^2\nn
\en
is independent on variable $\phi$ at $a_1=a_2$. If $a_k$ are constants it means that two diagonal elements of inertia tensor of the body $a_1^{-1}=a_2^{-1}$ are equal to each other and we discuss \textit{symmetric} rigid body \cite{bm05}.

Due to the special  form of $P'_T$ (\ref{p2-sph2}) and additional assumptions (\ref{pi2-sph}-\ref{assum2}), equations (\ref{w-eq}) decompose on the subsystem of equations for $c_{ij}^{km}(q)$, subsystem of equations for $\mathrm z_k(p),c_{ij}^{km}(q)$ and third subsystem of equations for $\mathrm x_k(q), \mathrm y_k(q), c_{ij}^{km}(q)$, which can be partially solved independently to each other.

\begin{prop}
If assumptions (\ref{pi2-sph}-\ref{assum2}) hold, then subsystem of equations for the functions $z_k(p)$ coming in (\ref{w-eq}) has three families of solutions
\bq\label{3-fam}\begin{array}{lll}
\mbox{Case }\,1.\quad& \Pi_{ij}=0; \\ \\
\mbox{Case }\,2.& \mathrm z_{1}=0\,,\quad &\mathrm z_2=0\,; \\ \\
\mbox{Case }\,3.& \mathrm z_{1}=\dfrac{p_{\phi}}{3}\,,& \mathrm z_{2}=\dfrac{p_{\theta}}{3}\,.
 \end{array}
\eq
\end{prop}
 This proposition gives only the necessary conditions. Of course, there remain complementary equations on the other functions $c_{ij}^{km}(\theta)$, $\mathrm x_{jk}(\theta)$ and $\mathrm y_{jk}(\theta)$ which have to be solved in the sequel.

At the first case $P'_T=0$ and we can immediately look for compatible potential part $\Lambda(\phi,\theta)$ and the variables of separation $u_{1,2}$ (\ref{dn-var}), which are related with initial variables by the point canonical transformations
\bq\label{point-trans}
u_i=f_i(\phi,\theta),\qquad p_{u_i}=g_i(\phi,\theta)\,p_\phi+h_i(\phi,\theta)\,p_\theta\,.
\eq
As a consequence, the geodesic Hamiltonian is a second order homogeneous polynomial in physical and separated momenta and the theory of projectively equivalent metrics in classical differential
geometry study essentially the same object \cite{bm03}.

\begin{prop}
In second case generic solution of (\ref{w-eq}) is parameterized by six functions $g,h$ and one parameter $\gamma=0,1$:
\bq\label{a-z}
\Pi=\left(
 \begin{array}{cc}
 \gamma p_\phi^2 & g_1(\theta)p_\phi^2+ g_2(\theta)p_\phi\,p_\theta+ g_3(\theta)p_\theta^2\\ \\
 0 & h_1(\theta)p_\phi^2+ h_2(\theta)p_\phi\,p_\theta+ h_3(\theta)p_\theta^2
 \end{array}
 \right)
\eq
up to the point transformations $p_k\to \alpha_k p_1+\beta_k p_2$.
\end{prop}
As above it is only necessary condition and functions $g,h$ from (\ref{a-z}), together with
functions $\mathrm x,\mathrm y$ from (\ref{p2-sph2}), are solutions of the remaining six non-linear differential
 equations in (\ref{w-eq}).
\begin{prop}
 In third case generic solution of
(\ref{w-eq}) is parameterized by nine functions $f,g,h$ and one parameter $\gamma=0,1$:
\bq\label{a-nz}
\Pi=\left(
 \begin{array}{cc}
 f_1(\theta)p_\phi^2+ f_2(\theta)p_\phi\,p_\theta+ f_3(\theta)p_\theta^2 &g_1(\theta)p_\phi^2+ g_2(\theta)p_\phi\,p_\theta+ g_3(\theta)p_\theta^2\\ \\
\frac12f_2(\theta)p_\phi^2+2f_3(\theta)p_\phi\,p_\theta+\gamma\Bigl(f_3(\theta)+h_3(\theta)\Bigr)^{3/2}p_\theta^2 & h_1(\theta)p_\phi^2+ h_2(\theta)p_\phi\,p_\theta+ h_3(\theta)p_\theta^2
 \end{array}
 \right)\,
\eq
up to the point transformations $p_k\to \alpha_k p_1+\beta_k p_2$.
\end{prop}
Functions $f,g,h$ from (\ref{a-nz}), together with functions $\mathrm x,\mathrm y$ from (\ref{p2-sph2}), are solutions of the remaining 19 non-linear differential equations in (\ref{w-eq}).

Matrices (\ref{a-z}) and (\ref{a-nz}) were obtained as solutions of the
subsystem of algebraic and linear differential equations for $c_{ij}^{km}(\theta)$, which has an unambiguous solution. The remaining functions satisfy to the complementary overdetermined subsystem of nonlinear PDE's, which have many distinct particular solutions.

In both cases (\ref{a-z}) and (\ref{a-nz}) we can get a complete classification of these particular solutions and of the corresponding bi-Hamiltonian systems (\ref{aux-int}).  Classification of separable bi-integrable systems demands additional assumptions on the form of the separated relations.

\subsection{Case 2 - classification of natural bi-Hamiltonian systems}
Let us briefly discuss a procedure of classification of the natural bi-Hamiltonian systems associated with natural Poisson bivector (\ref{n-p}-\ref{p2-sph2}) defined by the geodesic matrix $\Pi$ (\ref{a-z}).

If $h_2(\theta)=0$ in (\ref{a-z}), then six differential equations coming in (\ref{w-eq}) have four distinct solutions; among them we pick out solution defined by the following matrix
\[
\Pi=\left(
 \begin{array}{cc}
 \gamma\,p_\phi^2 &\gamma\left(1-\dfrac{h'_3(\theta)\,F}{\alpha\sqrt{h_3(\theta}}+F^2\right) p_\phi\,p_\theta \\ \\
 0 & \gamma\left(1+F^2\right)p_\phi^2+h_3(\theta)\,p_\theta^2 \\
 \end{array}
 \right)\,,\qquad F=\tan\left(\alpha\int{\dfrac{d\theta}{\sqrt{h_3(\theta)}}}+\beta\right)
\]
If $a_1=a_2=const$, then we can put $h_3(\theta)=\gamma=1$ without loss of generality and obtain
 \bq\label{nph-pi}
 \Pi=\left(
 \begin{array}{cc}
 p_\phi^2 & (1+\tan^2\alpha\theta)\,p_\phi\,p_\theta\\ \\
 0 & (1+\tan^2\alpha\theta)\,p_\phi^2+p_\theta^2
 \end{array}
 \right)\,,\qquad
\mathrm y_{12}(\theta)=\dfrac{2\alpha\mathrm x_{2 2}(\theta)-\cos\alpha\theta\sin\alpha\theta}{\alpha}\,.
\eq
The corresponding geodesic Hamiltonian (\ref{aux-int}) is equal to
\[
\mathcal T=\dfrac{1}{2}\,\mathrm{tr}\,N=\mathrm{tr}\,\Pi=(2+\tan^2\alpha\theta)p_\phi^2+p_\theta^2\,.
\]
At $\alpha=1$ matrix $\Pi$ (\ref{nph-pi}) is consistent only with the following potential matrix
\bq\label{nph-l}
\Lambda=\left(
 \begin{array}{cc}
 f(\phi) &g(\phi,\theta) \\ \\
 \dfrac{f'(\phi)\sin\theta}{2\cos\theta}-g(\phi,\theta)\quad & \dfrac{2\cos2\phi(2\cos^2\theta+1)g(\phi,\theta)}{\sin2\phi\sin2\theta}+\dfrac{f(\phi)}{\cos^2\theta}+a\tan^2\theta
 \end{array}
 \right)\,,
\eq
where
\ben
f(\phi)&=&a\cot^2\phi+\dfrac{b}{\sin^2\phi}+\dfrac{c}{\sin^2\phi\cos^2\phi}+\dfrac{2d\cos^2\phi(2\cos^2\phi-3)}{\sin^2\phi}\,,
\nn\\
g(\phi,\theta)&=& \dfrac{2d\sin^3\theta\sin2\phi}{\cos\theta}\,.\nn
\en
So, bi-Hamiltonian system associated with $\Pi$ (\ref{nph-pi}) and $\Lambda$ (\ref{nph-l}) has the following Hamilton function (\ref{aux-int})
\ben
\mathcal H_1&=&\mathcal T+\dfrac{a\bigl((x_1^2+x_2^2)-x_3^2(x_1^2-x_2^2)\bigr)}{x_1^2x_3^2}+\dfrac{(1+x_3^2)(x_1^2+x_2^2)\bigl(bx_2^2+c(x_1^2+x_2^2)\bigr)}{x_1^2x_2^2x_3^2}\nn\\
&-&\dfrac{2d(x_1^2+x_2^2+2x_1^2x_3^2)\bigl((x_1^2+x_2^2)-x_3^2(x_1^2-x_2^2)\bigr)}{(x_1^2+x_2^2)x_1^2x_3^2}\,.
\nn
\en
Second integral of motion $\mathcal H_2$ (\ref{aux-int}) is a fourth order polynomial in momenta. This integrable system, to the best of our knowledge, has not been considered in literature yet.

In similar manner we can get a complete classification of natural bi-Hamiltonian systems associated with matrices (\ref{a-z}) and (\ref{a-nz}).

\subsection{Case 3 - one possible generalization}
Non-invariant assumptions (\ref{pi2-sph},\ref{assum2}) depend on a choice of coordinate system and we miss a lot of another solutions of (\ref{m-eq1}), which may be interesting in applications.

One of the possible generalizations consists in the application of multiplicative separable functions in (\ref{pi2-sph}) \[c_{ij}^{km}(\phi,\theta)=a_{ij}^{km}(\phi)\,b_{ij}^{km}(\theta)\,,\]
and similar for $\mathrm{x,y}$. For instance, geodesic matrix
\bq\label{pi-lagr2}
\Pi= e^{2\ii \phi}\left(
 \begin{array}{cc}
 (\sin\theta\,p_\theta+\ii \cos\theta\,p_\phi)^2 & \dfrac{\alpha^2p_\phi (\sin\theta\,p_\theta+\ii \cos\theta\,p_\phi)}{\sin^3\,\theta} \\ \\
 0& 0
 \end{array}
 \right)\,,\qquad \ii=\sqrt{-1}\,,\quad \alpha\in\mathbb C\,,
\eq
gives rise to the natural Poisson bivector $P'$ at
\[
\mathrm y_{11} =- \dfrac{\ii}{2}\,,\qquad\mathrm z_{1}=\dfrac{p_{\phi}}{3}\,,\qquad \mathrm z_{2}=\dfrac{p_{\theta}}{3}\,.
\]
It is easy to prove that integrals of motion for the Lagrange top (\ref{int-lagr}) are in involution with respect to the corresponding Poisson bracket $\{.,.\}'$.

\begin{rem}
 Bivector $P'_T$ (\ref{p2-sph2}) associated with $\Pi$ (\ref{pi-lagr2}) has a natural counterpart on the generic symplectic leaves of the Lie algebra $e^*(3)$ at $(x,J)\neq 0$.
\end{rem}

\subsection{ Case 3 - three-dimensional sphere}

On the three and four dimensional spheres endowed with the standard spherical coordinates there are the same three families of solutions (\ref{3-fam}). It means that factor $1/3$ in (\ref{3-fam}) is independent on dimension of the sphere.

For instance, if $q=(\phi,\psi,,\theta)$ and $p=(p_\phi,p_\psi,p_\theta)$ are the standard spherical coordinates on $T^*\mathbb S^3$, then at $z_k=\dfrac{p_k}{3}$ matrices
\ben
\Pi_1&=&\left(
 \begin{array}{ccc}
 p_\phi^2 &2p_\phi p_\psi &\left ( 4-\dfrac{2ff''}{f'^2}\right)p_\phi p_\theta \\
 0 & p_\phi^2+fp_\psi^2+\dfrac{\alpha f^3}{f'^2}\, p_\theta^2\quad & \left(2-\dfrac{4ff''}{3f'^2}\right)f\,p_\psi p_\theta \\
 0 & \dfrac{2\alpha f^3}{f'^2}\,p_\psi p_\theta & p_\phi^2-\dfrac{f}{3}p_\psi^2+\dfrac{\alpha f^3}{f'^2}p_\theta^2
 \end{array}
 \right)\,,\qquad f=f(\theta)\,,\nn\\
 \label{pi-sph3}\\
\Pi_2&=&
\left(
  \begin{array}{ccc}
    p_\phi^2 & 2p_\phi p_\psi  &2 (e^{\alpha\psi}+1)\,p_\phi p_\theta \\
    0 & F\quad &2\beta e^{-\alpha\psi}(e^{\alpha\psi}+1)^2p_\psi p_\theta  \\
    0 &-2\gamma e^{-\alpha\psi}p_\psi p_\theta  &F-4\gamma(e^{-\alpha\psi}+1)p_\theta^2
  \end{array}
\right)\,,\nn
\en
where  $F=(e^{\alpha\psi}+1)p_\phi^2+\beta e^{-\alpha\psi}(e^{\alpha\psi}+1)^2\,p_\psi^2+\gamma e^{-\alpha\psi}\,p_\theta^2$,
  determine geodesic Poisson bivectors (\ref{p2-sph2}) and geodesic Hamiltonians (\ref{aux-int})
  \[
\mathcal T_1=3p_\phi^2+\dfrac{2f}{3}\,p_\psi^2+\dfrac{2\alpha f^3}{f'^2}\,p_\theta^2\,,\qquad
\mathcal T_2=(2e^{\alpha\psi}+3)p_\phi^2+\dfrac{2\beta(e^{\alpha\psi}+1)^2}{e^{\alpha\psi}}\,p_\psi^2
-2\gamma\left(2+\dfrac{1}{e^{\alpha\psi}}\right)\,p_\theta^2\,.
\]
Then we can calculate  compatible potential matrices $\Lambda_{1,2}$ depending on coordinates $(\phi,\psi,\theta)$ and the corresponding integrable potentials $V_{1,2}$. The corresponding integrals of motion $\mathcal H_{2,3}$ (\ref{aux-int}) are the fourth and sixth order polynomials in momenta, respectively.

So, using notion of the natural Poisson bivectors we can produce a lot of  abstract mathematical examples of bi-Hamiltonian system on the sphere. The main problems are how to select physically interesting bi-Hamiltonian systems and how to construct significant separable systems from the non-physical auxiliary bi-Hamiltonian systems.

\section{Separable bi-integrable systems }
\setcounter{equation}{0}
 In this Section we present matrices $\Pi$ and $\Lambda$ for the following well-known separable systems on the sphere
\begin{itemize}
 \item Case 1 - Lagrange top, Neumann system and systems separable in the elliptic coordinates;
 \item Case 2 - Goryachev system, Matveev-Dullin system, Kowalevsky top, Chaplygin system;
 \item Case 3 - Goryachev-Chaplygin top, Sokolov system, Kowalevsky\--Goryachev\--Chaplygin gyrostat;
\end{itemize}
which may be natively embedded into the proposed scheme as separable bi-integrable systems. Some new  mathematical generalizations of these systems and new separation of known systems are collateral results for this activity.

In framework of the Jacobi methods one gets integrals of motion $H_1,\ldots, H_n$ as solutions of the separated relations (\ref{seprel}). Of course, variables of separation and separated relations could have the singular points. So, the standard problem  is the rigorous determination of domain where variables of separation and integrals of motion are well defined, see the Jacobi definition of the elliptic coordinates.

Our main purpose is to discuss natural Poisson bivectors and, therefore, we do not comment this
 huge and complicated part of the work here, see for example \cite{bm05,chap04,dull04,gaff,kow89,yeh} and references within.

\subsection{Case 1 - Lagrange top}
If the spherical coordinates $\phi,\theta$ (\ref{sph-coord}) are variables of separation, one gets the simplest natural Poisson bivector $P'$ (\ref{p2-sph2}) at
\bq\label{p2-lagr}
 \Pi=0\,,\qquad \mbox{and}\qquad \Lambda=\left(
 \begin{array}{cc}
 \phi & 0 \\
 0 & \theta
 \end{array}
 \right)\,.
\eq
The auxiliary bi-Hamiltonian system is trivial
\[\mathcal H_1=\phi+\theta\,,\qquad \mathcal H_2=\dfrac12\,(\phi^2+\theta^2).\]
On the other hand, substituting variables of separation $u_1=\phi$ and $u_2=\theta$ into the separated relations
\[
\Phi_1=\left(a+\dfrac{\cos^2\theta}{\sin^2\theta}\right)H_2-H_1+p_{\theta}^2+b \cos\theta=0\,,\qquad \Phi_2=p_\phi^2-H_2=0\,,
\]
one gets integrals of motion for the Lagrange top in rotating frame
\bq\label{int-lagr}
H_1=J_1^2+J_2^2+aJ_3^2+bx_3\,,\qquad H_2=J_3^2\,, \qquad a,b\in \mathbb R\,,
\eq
More complicated natural bi-vector $P'$ obtained from matrix (\ref{pi-lagr2}) gives rise to another variables of separation for this system.
\begin{rem}
According to \cite{ts08} bivector $P'$ (\ref{n-p}) associated with $\Lambda$ (\ref{p2-lagr}) admits extension from cotangent bundle $T^*\mathbb S^2$ to the symplectic leaves of the Lie algebra $e^*(3)$ at $(x,J)\neq 0$.
\end{rem}

\subsection{ Case 1 - Neumann system }
Let us put $P'_T=0$ in (\ref{n-p}) and consider some particular solution $P'$ of the equations (\ref{m-eq1}) defined by the following non-symmetric matrix
\bq\label{p2-neu}
\Lambda=\left(
 \begin{array}{cc}
 a_1 \cos^2\phi+a_2\sin^2\phi & \dfrac{(a_1-a_2)\sin 2\phi}2\,\dfrac{\cos\theta}{\sin\theta}\\
 \\
 \dfrac{(a_1-a_2)\sin 2\phi}2\,{\cos\theta}\,{\sin\theta}\quad & a_3\sin^2\theta+(a_1\sin^2\phi+a_2\cos^2\phi)\cos^2\theta
 \end{array}
 \right)\,
\eq
with three arbitrary parameters $a_k\in\mathbb R$. As above, the auxiliary bi-Hamiltonian system has trivial integrals of motion $\mathcal H_{k}$ (\ref{aux-int}), which are functions only on the configurational space $\mathbb S^2$.

On the other hand, coordinates of separation $u_{j}$ (\ref{dn-var}) are the standard elliptic coordinates on the sphere
\bq\label{ell-q}
\dfrac{x_1^2}{\lambda-a_1}+\dfrac{x_2^2}{\lambda-a_2}+\dfrac{x_3^2}{\lambda-a_3}
=\dfrac{(\lambda-u_1)(\lambda-u_2)}{(\lambda-a_1)(\lambda-a_2)(\lambda-a_3)}\,.\qquad \eq
By substituting these variables in the separated relations
\[
u_i H_1-H_2-4(a_1-u_i)(a_2-u_i)(a_3-u_i)\,p_{u_i}^2+U_i(u_i)=0,\qquad i=1,2\,,
\]
one gets bi-integrable systems with quadratic in momenta integrals of motion
\[
H_1=J_1^2+J_2^2+J_3^2+V(x)\,,\qquad H_2=a_1J_1^2+a_2J_2^2+a_3J_3^2+W(x) \,,
\]
which are in the bi-involution (\ref{bi-inv}) with respect to both Poisson brackets.
Here $V(x)$ and $W(x)$ are easy calculated from the potentials $U_{1,2}$.
For instance, if
\[U(u)=u(u-a_1-a_2-a_3)\,,\]
then one gets the Neumann system with the following integrals of motion
\bq\label{h-neu}
 \begin{array}{l}
 H_1=J_1^2+J_2^2+J_3^2+a_1x_1+a_2x_2+a_3x_3\,, \\ \\
 H_2=a_1J_1^2+a_2J_2^2+a_3J_3^2-a_2a_3x_1-a_1a_3x_2-a_1a_2x_3 \,. \\
 \end{array}
\eq
\begin{rem}
Bivector $P'$ (\ref{n-p}) associated with $\Lambda$ (\ref{p2-neu}) also satisfies equations (\ref{m-eq1}) at $(x,J)\neq 0$, but in this case we lose bi-involutivity (\ref{bi-inv}) of integrals of motion $H_{1,2}$ (\ref{h-neu}) for the Clebsch system on the whole phase space $e^*(3)$. Of course, the corresponding elliptic coordinates on $e^*(3)$ remain variables of separation, but we can not get interesting natural Hamiltonians using these variables \cite{ts06}.
\end{rem}

\subsection{Case 2 - systems with cubic integral of motion }
At $\gamma=0$ in (\ref{a-z}) we have particular solution of the equations (\ref{w-eq}) defined by geodesic matrix
 \bq\label{gor-pi}
 \Pi=\left(
 \begin{array}{cc}
 0 & -\dfrac{\mathrm i}{2}\,\left(\dfrac{\partial }{\partial \theta}+\dfrac{2h(\theta)}{g(\theta)} \right)\,F\\ \\
 0 & F
 \end{array}
 \right)\,,\qquad F=\Bigl(g(\theta)p_\theta-\mathrm i h(\theta)p_\phi\Bigr)^2\,,\quad \mathrm i=\sqrt{-1}\,,
 \eq
depending on arbitrary functions $g(\theta)$ and $h(\theta)$ and by functions
\[
\mathrm x_{22}=-\dfrac{g(\theta)}{2h(\theta)}\,,\qquad \mathrm y_{12}=0,\qquad \mathrm z_k=0\,.
\]
This matrix $\Pi$ is consistent with the diagonal potential matrix
\bq\label{gor-l}
\Lambda=\alpha\,\exp\left(\mathrm i\phi-\int\dfrac{h(\theta)}{g(\theta)}\,d\theta\right)\left(
 \begin{array}{cc}
 1 & 0 \\ \\
 0 & 1
 \end{array}
 \right)\,.
\eq
The corresponding bi-Hamiltonian systems (\ref{aux-int}) are non-physical $\mathcal T=F$
 and, therefore, we immediately proceed to consideration of the coordinates of separation $v_{1,2}=\sqrt{\,u_{1,2}}$ following to \cite{ts09v}. If we introduce polynomial
\[
\mathcal B(\lambda)=(\lambda-v_1)(\lambda-v_2)=
\lambda^2-\mathrm i\sqrt{F}\lambda+\Lambda_{1,1}\,.
\]
instead of characteristic polynomial $B(\lambda)=(\lambda-u_1)(\lambda-u_2)=(\lambda-v_1^2)(\lambda-v_2^2)$ (\ref{dn-var}) of recursion operator $N$, then it is easy to prove that
\[
\{\mathcal B(\lambda),\mathcal A(\mu)\}=\dfrac{\lambda}{\mu-\lambda}
\left(\dfrac{\mathcal B(\lambda)}{\lambda}-\dfrac{\mathcal B(\mu)}{\mu}\right)\,,
\qquad \{\mathcal A(\lambda),\mathcal A(\mu)\}=0\,,\]
where
\[
\mathcal A(\lambda)=\int\dfrac{\mathrm id\theta}{{g(\theta)}}-\dfrac{\mathrm ip_\phi}{\lambda}\,.
\]
It entails that
\[
p_{v_j}=\mathcal A(\lambda=v_j)\,,\qquad j=1,2,
\]
are canonically conjugated to $v_j$ momenta and that the corresponding Poisson brackets read as
\[
\{v_i,p_{v_j}\}=\delta_{ij},\qquad \{v_i,p_{v_j}\}'=\delta_{ij} v_i^2\,.
\]
Now we have to substitute this family of variables of separation into the separated relations and try to get natural Hamiltonians.
For instance, let us take
\bq\label{hg-gor}
g(\theta)=\sin\theta\,f(\theta)\,,\qquad h(\theta)=\cos\theta\, f(\theta)\,,
\eq
substitute
\[\lambda=v_j\qquad \mu=\dfrac{2\mathrm i}{3}\, v_{j}\,p_{j},\qquad j=1,2,\]
into the equation
\bq\label{gor-eq}
\Phi(\lambda,\mu)=\mu H_1+H_2-\mu^3-\lambda^3-b\lambda+\dfrac{\alpha^2}{\lambda}=0,
\eq
and solve a pair of the resulting equations with respect to $H_{1,2}$. If $a_1=a_2$ in the geodesic Hamiltonian (\ref{n-hams}), then in this solution we have to put
\[f(\theta)=\dfrac{\cos^{1/3}\theta}{\sin^2\theta} \,, \]
and we obtain integrals of motion for the Gorychev system
on the sphere \cite{gor15}
\ben
H_1&=&J_1^2+J_2^2+\dfrac{4}{3}J_3^2+\dfrac{2\mathrm i\alpha x_1}{x_3^{2/3}}-\dfrac{b}{x_3^{2/3}}\,,
\nn\\
\label{gor-int}\\
H_2&=&\dfrac{2J_3}{3}\left(J_1^2+J_2^2+\dfrac{8}{9}J_3^2-\dfrac{b}{x_3^{2/3}}\right)-2\mathrm i\alpha x_3^{1/3}J_1+\dfrac{4\mathrm i\alpha}{3x_3^{2/3}}x_1J_3\,.
\nn
\en
For other separable natural bi-integrable systems from \cite{ts05,ts09v} we present Hamiltonians and functions $g$ and $h$ only. So, for the Goryachev-Chaplygin top \cite{gor00,chap04} we have
\[
H_1=J_1^2+J_2^2+4J_3^2+ax_1+\dfrac{b}{x_3^2}\,,\qquad g(\theta) = \dfrac{1}{\cos\theta\sin\theta}\,,\qquad h(\theta) =\dfrac{3\cos^2\theta-2}{\cos^2\theta\sin^2\theta}\,.\]
For the Dullin-Matveev system \cite{dull04} with Hamiltonian
\[H_1=J_1^2+J_2^2+\left(1+\dfrac{x_3}{x_3+c}-\dfrac{x_3^2-|x|^2}{4(x_3+c)^2}\right)
J_3^2+\dfrac{ax_1}{(x_3+c)^{1/2}}+\dfrac{b}{x_3+c}\]
geodesic matrix $\Pi$ (\ref{gor-pi}) and potential matrix $\Lambda$ (\ref{gor-l}) are defined by functions
\[
g(\theta) = \dfrac{1}{\sin\theta}\,,\qquad \qquad h(\theta)=-\dfrac{1-2c\cos\theta-3\cos^2\theta}{2\sin^2\theta\,(\cos\theta+c)}\,.
\]
For the system with the Hamiltonian 
\[
H_1=J_1^2+J_2^2+\left(\dfrac{7}{12}+\dfrac{x_3}{2(x_3+|x|)}\right)
J_3^2
+\dfrac{2\mathrm i\alpha x_1}{(x_3+|x|)^{5/6}}-\dfrac{b}{(x_3+|x|)^{1/3}}\,,
\]
bi-Hamiltonian structure is defined by functions
\[
g(\theta) = \dfrac{(\cos\theta+1)^{2/3}}{\sin\theta}\,,\qquad\qquad
 h(\theta) = -\dfrac{(\cos\theta+1)^{2/3}}{2(\cos\theta)-1)}\,.
\]
For the last system from \cite{ts05} we have 
\[
H=J_1^2+J_2^2+\left(\dfrac{13}{16}+\dfrac{3x_3}{8(x_3+|x|)}\right)
J_3^2
+\dfrac{ax_1}{(x_3+|x|)^{3/4}}+\dfrac{b}{(x_3+|x|)^{1/2}}
\]
and
 \[
 g(\theta) =\dfrac{ (\cos\theta+1)^{1/2}}{\sin\theta}\,,\qquad\qquad
h(\theta)=\dfrac{ (3\cos\theta+1)(\cos\theta+1)^{1/2}}{4\sin^2\theta}\,.
 \]
 If $\widetilde{P}'$ is the linear in momenta Poisson bivector from \cite{ts09v}, then
our natural Poisson bivector is equal to $P'=\widetilde{P}'P^{-1}\widetilde{P}'$.

\begin{rem}
According to \cite{val10} , the Coryachev-Chaplygin, Chaplygin and Dullin-Matveev systems can be embedded into a family of integrable systems with cubic integral of motion. We suppose that bi-Hamiltonian structures for the Valent systems may be described by a suitable choice of the functions $g(\theta)$ and $h(\theta)$ in (\ref{gor-pi}) and (\ref{gor-l}).
\end{rem}

\begin{rem}
Another possible generalization consists of multiplication of matrix $\Pi$ (\ref{gor-pi}) on the functions depending on $\phi$ similar to (\ref{pi-lagr2}).
\end{rem}

\subsection{Case 2 - Kowalevski top and Chaplygin system}
 Let us consider a geodesic bivector $P'_T$ (\ref{p2-sph2}) determined by the matrix $\Pi$
\bq\label{kow-pi}
\Pi=\dfrac{1}{\sin^\alpha \theta\, \cos^2\theta}\left(
 \begin{array}{cc}
 0 & \dfrac{2\,p_\phi\,p_\theta}{\alpha} \\ \\
 0 & \cos^2\theta\,p_\phi^2+\sin^2\theta\,p_\theta^2
 \end{array}
 \right)\,,\qquad \alpha\in \mathbb R\,,
\eq
and by functions
\[\mathrm y_{12}(\theta)=\cos\theta\Bigl(\sin\theta+\alpha \mathrm x_{22}(\theta)\cos\theta\Bigr)\,,\qquad \mathrm z_{1,2}=0\,.
\]
There is only one potential matrix consistent with
 $\Pi$ (\ref{kow-pi})
\bq\label{kow-l}
\Lambda=\left(
 \begin{array}{cc}
 a\cos \alpha\phi-b\sin \alpha\phi & \Bigl (a\sin \alpha\phi-b\cos \alpha\phi\Bigr)\cot\theta \\ \\
 \Bigl (a\sin \alpha\phi-b\cos \alpha\phi\Bigr) \tan\theta & -a\cos \alpha\phi+b\sin \alpha\phi
 \end{array}
 \right)\,,\qquad a,b\in\mathbb R\,.
\eq
The corresponding coordinates of separation $u_{1,2}$ (\ref{dn-var}) are the roots of the polynomial
\ben
B(\lambda)&=&\lambda^2-\dfrac{p_\theta^2\sin^2\theta+p_\phi^2\cos^2\theta}{\sin^\alpha\theta\cos^2\theta}\,\lambda
-\dfrac{(a\cos\alpha\phi-b\sin\alpha\phi)(p_\theta^2\sin^2\theta+p_\phi^2\cos^2\theta)}
{\sin^\alpha\theta\cos^2\theta}\nn\\
\label{kow-var}\\
&-&\dfrac{2\sin\theta(a\sin\alpha\phi+b\cos\alpha\phi)p_\phi\,p_\theta}{\sin^\alpha\theta\cos^2\theta}-a^2-b^2\,.\nn
\en
Following to \cite{ts10a,ts10k} we can introduce auxiliary polynomial
\[
 A(\lambda)=\dfrac{\sin\theta p_\theta}{\alpha\cos\theta}\,\lambda
+\dfrac{a\sin\alpha\phi+b\cos\alpha\phi}{\alpha}\,p_\phi - \dfrac{\sin\theta(a\cos\alpha\phi-b\sin\alpha\phi)}{\alpha\cos\theta}\,p_\theta\,,
\]
such as
\[
\{B(\lambda), A(\mu)\}=\dfrac{1}{\lambda-\mu}\,\Bigl((\mu^2-a^2-b^2)B(\lambda)-(\lambda^2-a^2-b^2)B(\mu)\Bigr)\,,\qquad \{A(\lambda),A(\mu)\}=0\,.
\]
It entails that
\[
p_{u_j}=-\dfrac{1}{u_j^2-a^2-b^2}\, A(\lambda=u_j)\,,\qquad j=1,2,
\]
are the canonically conjugated momenta satisfying to the Poisson brackets (\ref{poi-br12}). At $\alpha=2$ these variables have been considered by Chaplygin \cite{chap03}.

By substituting these variables of separation into a pair of the separated relations
\[
\Phi_1=(u_1^2-a^2-b^2)p_{u_1}^2+H_1-H_2=0\,,\qquad \Phi_2= (u_2^2-a^2-b^2)p_{u_2}^2+H_1+H_2=0\,,
\]
one gets separable bi-integrable system with the Hamilton function
\bq\label{g-kgch}
2\alpha^2\,H_1=p_\phi^2-\tan^2\theta\, p_\theta^2+2(a\cos\alpha\phi+b\cos\alpha\phi)\,\cos^\alpha\theta\,,\qquad \alpha\in\mathbb R\,.
\eq
According to \cite{ts10a,ts10k}, at $\alpha=1$ using separated relations
\bq\label{kow-sep}
\Phi(u,p_u)=\Bigl((u^2-a^2-b^2)p_u^2+H_1-H_2\Bigr)\Bigl((u^2-a^2-b^2)p_u+H_1+H_2\Bigr)+cu^2+du=0
\eq
one gets Hamilton function of the generalized Kowalevski top \cite{kow89}
\bq\label{kow-hg}
H^{kow}=2H_1=\left(1-\dfrac{c+1}{x_3^2}\right)(J_1^2+J_2^2)+2J_3^2+2ax_2+2bx_1-\dfrac{d}{\sqrt{x_1^2+x_2^2}}\,.
\eq
At $\alpha=2$ we can use another separated relations
\bq\label{chap-sep}
\Phi(u,p_u)=\Bigl((u^2-a^2-b^2)p_u^2+cu+H_1-H_2\Bigr)\Bigl((u^2-a^2-b^2)p_u^2+cu+H_1+H_2)+du=0
\eq
in order to get Hamiltonian of the generalized Chaplygin system \cite{chap03,gor16}
\bq\label{chap-hg}
H^{ch}=8H_1=\left(1-\dfrac{4c+1}{x_3^2}\right)(J_1^2+J_2^2)+2J_3^2-2a(x_1^2-x_2^2)-2bx_1x_2-\dfrac{2d}{1+4c-x_3^2}\,.
\eq
At $c=-\alpha^{-2}$ we have geodesic Hamiltonian $T=J_1^2+J_2^2+2J_3^2$ with the constant inertia tensor.

\begin{rem}
By substituting these variables of separation into another separation relations we
can obtain various mathematical generalizations of bi-integrable Hamiltonians (\ref{g-kgch},\ref{kow-hg},\ref{chap-hg}).
\end{rem}

 \subsection{Case 2 - spherical top and Chaplygin system}
At $\gamma=0$ in (\ref{a-z}) we have a particular solution of the equations (\ref{w-eq}) defined by matrix
 \bq\label{p-tr-ex}
\Pi=\left(
 \begin{array}{cc}
 p_\phi^2 & \dfrac{\alpha-\sin^2\theta}{\cos^2\theta\sin^2\theta}\,p_\phi\, p_\theta \\ \\
 0 & \dfrac{\alpha}{\sin^2\theta}\,p_\phi^2+\dfrac{\alpha-\sin^2\theta}{\cos^2\theta}\,p_\theta^2
 \end{array}
 \right)\,,\qquad \alpha\in\mathbb R\,,
\eq
and functions
 \[
 \mathrm y_{12}= \sin\theta\cos\theta+\dfrac{2\alpha\cos^2\theta}{\sin^2\theta-\alpha}\,\mathrm x_{22}\,,\qquad \mathrm z_k=0\,.
 \]
In this case coordinates of separation $u_{1,2}$ (\ref{dn-var}) are equal to
\[
u_1=p_\phi^2\,,\qquad\qquad u_2=\dfrac{\alpha p_\phi^2}{\sin^2\theta}-\dfrac{(\sin^2\theta-\alpha)p_\theta^2}{\cos^2\theta}\,,
\]
so that conjugated momenta read as
\[
p_{u_1}=\dfrac{\arctan\left(\dfrac{p_\theta\tan\theta}{p_\phi}\right)-\phi}{2p_\phi}\,,\qquad p_{u_2}=\dfrac{\mathrm \sin\theta\cos\theta\arctan\left(\dfrac{\sin^2\theta\,p_\theta}{\sqrt{\alpha\cos^2\theta p_\phi^2-\sin^2\theta(\sin^2\theta-\alpha)p_\theta^2}}\right)}
{2\sqrt{\alpha\cos^2\theta p_\phi^2-\sin^2\theta(\sin^2\theta-\alpha)p_\theta^2}}\,.
\]
By substituting these variables of separation into the separated relations
\[
\Phi_1=\sqrt{u_1}-H_2=0\,,\qquad\Phi_2=\alpha H_1-u_2\Bigl(1-(\alpha-1)\tan^2(2p_{u_2}\sqrt{u_2})\Bigr)+\alpha f(\theta)=0\,,
\]
where
\[
\theta=\arccos\sqrt{\dfrac{u_2-\alpha H_2^2}{u_2}+\dfrac{\alpha(H_2^2-u_2)(1-\cos4p_{u_2}\sqrt{u_2})}{2u_2}}
\]
one gets generalized Lagrange top with integrals of motion
\bq\label{eq-glag}
H_1=J_1^2+J_2^2+J_3^2+f(x_3)\,,\qquad H_2=J_3\,.
\eq
Other separated relations
\ben
\Phi_1(u_1,p_{u_1})&=&\dfrac{2}{\sqrt{u_1}\,\sin\left(4p_{u_1}\sqrt{u_1}\right)}\,H_2-H_1+u_1=0\,,
\nn\\
\label{seprel-sph}\\
\Phi_2(u_1,p_{u_1})&=&\alpha H_1-u_2\Bigl(1-(\alpha-1)\tan^2(2p_{u_2}\sqrt{u_2})\Bigr)=0\,.\nn
\en
give rise to integrals of motion for the spherical top
\bq\label{sph-top}
 H_1=T=J_1^2+J_2^2+J_3^2\,,\qquad \qquad H_2=J_1J_2J_3\,.
\eq
 There are only two potential matrices compatible with $\Pi$ (\ref{p-tr-ex})
\[
\Lambda^{(1)}=\left(
 \begin{array}{cc}
 f(\phi) & 0 \\ \\
 \dfrac{f'(\phi)(\sin^2\theta-\alpha)}{2\sin\theta\cos\theta} &\dfrac{ \alpha f(\phi) }{\sin^2\theta}
 \end{array}
 \right)\]
and
\[
\Lambda^{(2)}=\left(
 \begin{array}{cc}
 a \sin2\phi+b \cos2\phi & -\dfrac{\cos\theta}{\alpha\sin\theta}(\alpha-\sin^2\theta)(a\cos2\phi-b \sin2\phi) \\ \\
 -\dfrac{\sin\theta}{\alpha\cos\theta}(\alpha-\sin^2\theta)(a\cos2\phi-b \sin2\phi) &
-\dfrac{(\alpha-2\sin^2\theta)(a\sin2\phi+b \cos2\phi)}{\alpha}
 \end{array}
 \right)\,.
\]
In the first case the auxiliary bi-Hamiltonian system with the Hamilton function (\ref{aux-int})
\bq\label{def-kow1}
\mathcal H_1^{(1)}=\left(1+\dfrac{\alpha-1}{x_3^2}\right)\Bigl(J_1^2+J_2^2\Bigr)+2J_3^2
+f\left(\dfrac{x_1}{x_2}\right)\left(1+\dfrac{\alpha}{x_1^2+x_2^2}\right)\,,
\eq
is a deformation of the geodesic Hamiltonian for the Kowalevski top at $\alpha=1$ and $f=0$. By substituting the corresponding coordinates of separation (\ref{dn-var})
\[
\hat{u}_1=u_1+f(\phi)\,,
\quad\hat{p}_{u_1}=p_{u_1}-\dfrac{1}{2}\int^\phi\dfrac{dx}{p_\phi^2+f(\phi)-f(x)}\,,
\qquad \hat{u}_2=u_2+\dfrac{\alpha f(\phi)}{\sin^2\theta}\,,
\quad \hat{p}_{u_2}=p_{u_2}\,,
\]
into $\Phi_1=\hat{u}_1-\widehat{H}_2=0 $ and the second separated relation $\Phi_2$ in (\ref{seprel-sph}), one gets a generalization of the spherical top defined by the following integrals of motion
\[
\widehat{H}_1=J_1^2+J_2^2+J_3^2+\dfrac{f\left(\dfrac{x_1}{x_2}\right)}{x_1^2+x_2^2}\,,\qquad
\widehat{H}_2=J_3^2+\dfrac{f\left(\dfrac{x_1}{x_2}\right)}{x_1^2+x_2^2+x_3^2}.
\]
In the second case matrices $\Pi$ (\ref{p-tr-ex}) and $\Lambda^{(2)}$ give rise to the auxiliary bi-Hamiltonian system with the Hamilton function
\bq\label{def-chap1}
\mathcal H_1^{(2)}=\left(1+\dfrac{\alpha-1}{x_3^2}\right)\Bigl(J_1^2+J_2^2\Bigr)+2J_3^2+
4\alpha^{-1}ax_1x_2-2\alpha^{-1}b(x_1^2-x_2^2)\,.
\eq
It is a new deformation of the well-known Chaplygin system \cite{chap03}.

\begin{rem}
According to \cite{ts99b}, there is a non-canonical map, which relates integrals of motion (\ref{sph-top}) with integrals of motion for the Gaffet system \cite{gaff}
\[
H_1=J_1^2+J_2^2+J_3^2-\dfrac{1}{(x_1 x_2 x_3)^{2/3}}\,,\qquad
H_2=J_1J_2J_3+\dfrac{x_2 x_3 J_1+x_1 x_3 J_2+x_1 x_2 J_3}{ (x_1 x_2 x_3)^{2/3} }\,.
\]
In order to describe the bi-Hamiltonian structure for the Gaffet system we have to use additional non-point transformation of the standard spherical coordinates, which changes the form of  $P'_T$ (\ref{p2-sph2}) in initial variables. This bi-Hamiltonian structure will be discussed in the forthcoming publication.
\end{rem}

\subsection{Case 3 - Goryachev-Chaplygin top and Sokolov system}
At $\gamma=0$ in (\ref{a-nz}) equations (\ref{w-eq}) have a particular solution $P'_T$ (\ref{p2-sph2}) defined by the following symmetric matrix
\bq\label{p-gch1}
\Pi=\left(
 \begin{array}{cc}
 p_\theta^2 +p_\phi^2(4+3\cot^2\alpha\theta) & 2p_\phi\,p_\theta \\ \\
 2p_\phi\,p_\theta &p_\theta^2-p_\phi^2\,\cot^2\alpha\theta
 \end{array}
 \right)\,,\qquad\alpha\in \mathbb R\,,
\eq
 and by the functions
\[
\mathrm x_{22}=\mathrm y_{12}=-\dfrac{\cos\alpha\theta\,\sin\alpha\theta}{\alpha}\,,\qquad\qquad \mathrm z_{k}=\dfrac{p_k}{3}\,.
\]
There is only one potential matrix compatible with $\Pi$ (\ref{p-gch1})
\[
\Lambda= \dfrac{a}{\cos^2\alpha\theta}\left(
 \begin{array}{cc}
 1 &0 \\
 0 & 1
 \end{array}
 \right)\,.
\]
The corresponding auxiliary bi-Hamiltonian system is defined by the Hamilton function (\ref{aux-int})
\[
\dfrac{1}{2}\mathcal H_1=(2+\cot^2\alpha\theta)p_\phi^2+p_\theta^2+\dfrac{a}{\cos^2\alpha\theta}\,.
\]
If $\alpha=1$, we have a deformation of the geodesic Hamiltonian for the Kowalevski top \cite{kow89}
\bq
\dfrac{1}{2}\,\mathcal H_1=J_1^2+J_2^2+2J_3^2+\dfrac{a}{x_3^2}\,.
\eq
This auxiliary bi-Hamiltonian system gives rise to the variables of separation $u_{1,2}$ (\ref{dn-var})
\[
u_{1,2}=\left(p_\phi\pm\sqrt{\dfrac{p_\phi^2}{\sin^2\theta}+p_\theta^2+\dfrac{a}{\cos^2\theta}}\right)^2=
\left(
J_3\pm\sqrt{J_1^2+J_2^2+J_3^2+\dfrac{a}{x_3^2}}
\right)^2\,,\qquad \alpha=1.
\]
At $a=0$ these coordinates were found in \cite{chap04}. By substituting
the generalized Chaplygin variables
\bq\label{fch-var}
v_{1,2}=\sqrt{u_{1,2}\,}\,,\qquad p_{v_{1,2}}=-\dfrac{1}{2\mathrm i}
\ln\Bigl(v_{1,2}(\mathrm i x_1-x_2)-(\mathrm i J_1-J_2)x_3\Bigr)+\dfrac{\ln(v_{1,2}^2-a)}{4\mathrm i}\,,
\eq
into the separated relations
\[
\Phi_{1,2}(v,p_v)=H_1v+H_2+b\sqrt{v^2-a}\sin 2 p_v-v^3-cv^2=0\,,\qquad v=v_{1,2},\quad p_v=p_{v_{1,2}}\,,
\]
 one gets integrals of motion for the generalized Goryachev-Chaplygin gyrostat \cite{gor00,chap04}
\ben
H_1&=&J_1^2+J_2^2+4J_3^2+2cJ_3+bx_1+\dfrac{a}{x_3^2}\,\nn\\
\nn\\
H_2&=&(2J_3+c)\left(J_1^2+J_2^2+\dfrac{a}{x_3^2}\right)-bx_3J_1\,
.\nn
\en
By substituting the same variables (\ref{fch-var}) into the following separated relations
\bq\label{sok-rel}
\Phi_{1,2}(v_{1,2},p_{v_{1,2}})=\widehat{H}_1\pm \widehat{H}_2+b\,\sqrt{v_{1,2}^2-a}\,\sin2p_{v_{1,2}}-v_{1,2}^2-cv_{1,2}=0\,,
\eq
we obtain the generalized Sokolov system \cite{sok} defined by integrals of motion
\ben
\widehat{H}_1&=&J_1^2+J_2^2+2J_3^2+c J_3+b(J_3x_1-x_3J_1)+\dfrac{a}{x_3^2}\,,\nn\\
\nn\\
\widehat{H}_2&=&\Bigl(2J_3+c+bx_1\Bigr)\sqrt{J_1^2+J_2^2+J_3^2+\dfrac{a}{x_3^2}}\,,\nn
\en
up to the canonical transformation discussed in \cite{kst03}.

\subsection{Case 3 - Kowalevski-Goryachev-Chaplygin gyrostat}
The geodesic matrix $\Pi$ (\ref{p-gch1}) for the Goryachev-Chaplygin top may be deformed
\bq\label{p-kowg}
\widehat{\Pi}=\Pi +\beta\left(
 \begin{array}{cc}
 0 & \dfrac{\cos\alpha\theta}{\sin^3\alpha\theta}\,p_\phi^2 \\ \\
 0 & 0
 \end{array}
 \right)\,,
\eq
if
\[
\mathrm y_{11}(\theta)=\mathrm x_{21}(\theta)-\dfrac{\beta}{2\alpha}\,,\qquad
y_{12}(\theta)=-\dfrac{\cos^2\alpha\theta}{\sin^2\alpha\theta}\,\mathrm x_{22}(\theta)
-\dfrac{\cos\alpha\theta}{\alpha \sin\alpha\theta}\,,\qquad \mathrm z_{k}=\dfrac{p_k}{3}.
\]

\begin{rem}
In the $r$-matrix formalism transition from matrix (\ref{p-gch1}) to the matrix (\ref{p-kowg}) generates transition from the quadratic Sklyanin bracket to the so-called reflection equation algebra \cite{ts02,ts08t}.
\end{rem}
In generic case matrix $\widehat{\Pi}$ (\ref{p-kowg}) is compatible with the potential matrix
\bq\label{lg-kowg}
\widehat{\Lambda}^{(1)}=
\dfrac{a\,\e^{-\frac{4\alpha\phi}{\beta}}}{\sin^2\alpha\theta}
\left(
 \begin{array}{cc}
 \cos^2\alpha\theta-4 & -\frac{\beta\cos\alpha\theta}{\sin\alpha\theta} \\ \\
 \frac{4\cos\alpha\theta\,\sin\alpha\theta}{\beta} & \cos^2\alpha\theta
 \end{array}
\right)+\frac{b\sin^2\alpha\theta}{\cos^2\alpha\theta}
\left(
 \begin{array}{cc}
 1 & -\frac{\beta}{\sin\alpha\theta\,\cos\alpha\theta} \\ \\
 0 & 1
 \end{array}
\right)\,
\eq
The corresponding auxiliary bi-Hamiltonian system is defined by the Hamiltonian
\[
\dfrac{1}{2}\mathcal H_1=(2+\cot^2\alpha\theta)p_\phi^2+p_\theta^2
+\dfrac{a(\cos^2\alpha\theta-2)\e^{-\frac{4\alpha\phi}{\beta}}}{\sin^2\alpha\theta}+\dfrac{b\sin^2\alpha\theta}{\cos^2\alpha\theta}\,.
\]
So, at $\alpha=1$ we have another deformation of the geodesic Hamiltonian for the Kowalevski top \cite{kow89}
\[
\dfrac{1}{2}\mathcal H_1=J_1^2+J_2^2+2J_3^2-\dfrac{ a(x_1^2+x_2^2+1)}{x_1^2+x_2^2}\,\e^{-\frac{4\arctan(x_1/x_2)}{\beta}}+\dfrac{b(x_1^2+x_2^2)}{x_3^2}\,.
\]
In this case description of the variables of separation
and the corresponding bi-integrable system is an open problem.

At $\beta=\pm2\,\mathrm i$ there is one more particular potential matrix compatible with $\widehat{\Pi}$ (\ref{p-kowg})
\bq\label{l-kowg}
\widehat{\Lambda}^{(2)}=\gamma\, \e^{\pm\mathrm i\alpha\phi}
\left(
 \begin{array}{cc}
 \pm\,\mathrm i\, \sin\alpha\theta & \cos\alpha\theta \\ \\
 0 & 0
 \end{array}
 \right)\,.
\eq
In this particular case we can substitute
the coordinates of separation $u_{1,2}$ (\ref{dn-var}) and the corresponding momenta $p_{u_{1,2}}$ into the separated relations defined by
\bq\label{kgyr-curve}
{\Phi}(u,p_u)=
u^6+H_1u^4+H_2u^2+a+\sqrt{b(u)}\sin 2p_u=0\,,
\eq
and obtain integrals of motion for the Kowalevski-Goryachev-Chaplygin gyrostat with the following Hamilton function
\bq\label{kgyr-h}
 H_1=J_1^2+J_2^2+2J_3^2+2c_1 J_3+c_2x_1+c_3(x_1^2-x_2^2)
+\dfrac{c_4}{x_3^2}\,,
\eq
see \cite{chap03,gor16,kow89,yeh}. Here $b(u)$ (\ref{kgyr-curve}) is a special polynomial of eight order in $u$ with coefficients depending on $a$ and $c_k$, see details in \cite{ts02}.

\begin{rem}
In this case in order to get the conjugated momenta $p_{u_{1,2}}$ and the separated relation we used the Lax matrices and the reflection equation algebra,
that drastically simplified all the calculations.
\end{rem}

\begin{rem}
For the systems with quartic integral of motion from \cite{ts05a} the natural Poisson bivector may be obtained using deformation of the matrix (\ref{p-kowg}) similar to (\ref{pi-lagr2}).
\end{rem}

\subsection{Case 2 - deformations of the Kowalevski top and Chaplygin systems}
 Let us consider trivial canonical transformation
 \bq\label{ptheta-tr}
 p_\theta\to p_\theta+f(\theta)\,,
 \eq
 which preserved canonical Poisson bivector $P$ (\ref{poi-0}). This mapping shifts the natural Poisson bivector $P'$ (\ref{n-p}) associated with matrices $\Pi$ (\ref{kow-pi}) and $\Lambda$ (\ref{kow-l}) by the rule
\[
\widehat{P}'=P'+g(\theta)\left(
 \begin{array}{cccc}
 0 & 0 & 0 &\left(\dfrac{\alpha\cos^2\theta-1}{\alpha\sin^2\theta}+\dfrac{\cot\theta}{\alpha}\,\ln'g\right) \,p_\phi \\ \\
 * & 0 & 0 &p_\theta+\dfrac{\cos^2\theta\,\sin^{\alpha-2}\theta}{4}\,g(\theta) \\ \\
 * & * & 0 & \dfrac{\sin^{\alpha-2}\theta(a\sin\alpha\phi+b\cos\alpha\phi)}{2}\Bigl(\sin\theta\cos\theta \ln'g-1\Bigr)\\ \\
 * & * & * & 0
 \end{array}
 \right)\,,
\]
where
\[g(\theta)=-\dfrac{2f(\theta)\sin^{2-\alpha}\theta}{\cos^2\theta}\,
\qquad\mbox{and}\qquad \ln'g=\dfrac{1}{g(\theta)}\dfrac{dg(\theta)}{d\theta}\,.
\]
The Poisson bivector $\widehat{P}'$ gives rise to the "shifted" variables of separation
\bq\label{sh-sepvar}
\hat{u}=\left. u\right|_{p_\theta\to p_\theta+f(\theta)}\,,\qquad
\hat{p_u}=\left. p_u \right|_{p_\theta\to p_\theta+f(\theta)}\,.
\eq
If we substitute these variables of separation into the old separated relations (\ref{kow-sep}) and (\ref{chap-sep})
one gets non-natural Hamiltonians, which are related to the old Hamiltonians (\ref{kow-hg}) and (\ref{chap-hg}) by canonical transformation (\ref{ptheta-tr}).

In order to get new natural Hamiltonians we have to appropriately modify the separated relations.
 For instance, let us take
\[ f(\theta)=\dfrac{\sqrt{\beta}\,\tan^{\alpha-1}\theta}{\cos^\alpha\theta}\,.\]
At $\alpha=1$ by substituting variables of separation (\ref{sh-sepvar}) into the new separated relations
\bq\label{dkow-sep}
\widehat{\Phi}=\Phi-\beta H_1+{\beta^2}+\sqrt{\beta}(\hat{u}^2-a^2-b^2)\hat{p}_{u},\qquad \hat{u}=\hat{u}_{1,2}\,,\quad
\hat{p}_u=\hat{p}_{u_{1,2}}\,,
\eq
where $\Phi$ is given by (\ref{kow-sep}), one gets generalization of the Hamilton function (\ref{kow-hg})
\[
\widehat{H}^{kow}=\left(1-\dfrac{c+1}{x_3^2}\right)(J_1^2+J_2^2)+2J_3^2+2ax_2+2bx_1-\dfrac{d}{\sqrt{x_1^2+x_2^2}}-\dfrac{\beta}{x_3}\,,\nn\\
\]
At $\alpha=2$ the "shifted" separated relations
\bq\label{dchap-sep}
\widehat{\Phi}=\Phi+\sqrt{\beta}(\hat{u}^2-a^2-b^2)\hat{p}_u\,,\qquad \hat{u}=\hat{u}_{1,2}\,,\quad
\hat{p}_u=\hat{p}_{u_{1,2}}\,,
\eq
where $\Phi$ is given by (\ref{chap-sep}), yield similar generalization of the Hamiltonian (\ref{chap-hg})
\[
\widehat{H}^{ch}=\left(1-\dfrac{4c+1}{x_3^2}\right)(J_1^2+J_2^2)+2J_3^2-2a(x_1^2-x_2^2)-2bx_1x_2-\dfrac{2d}{1+4c-x_3^2}+\beta\left(\dfrac{1}{x_3^4}-\dfrac{1}{x_3^6}\right)\,.\nn
\]
These Hamiltonians at $c=-\alpha^{-2}$ and another Hamiltonians associated with various functions $f(\theta)$ may be found in \cite{yeh}. 

The separability of these systems, to the best of our knowledge, has not been considered in literature yet. In tboth cases equations of motion are linearized on  the two copies of the non-hyperelliptic curves of genus three defined by  (\ref{dkow-sep}) and (\ref{dchap-sep}). We do not know how to solve the corresponding Abel-Jacobi equations as yet.

\begin{rem}
Other natural Poisson bivectors studied in the previous Sections may be shifted on the similar linear in momenta terms. As above, it allows us to get various generalizations of the considered bi-integrable systems.
\end{rem}

\section{Conclusion}
We proved that almost all known integrable systems on the two-dimensional unit sphere $\mathbb S$
 may be studied in the framework of a single theory of natural Poisson bivectors.
  It is an experimental fact supported by all the know constructions of the variables of separation on the sphere. We try to draw attention to this experimental fact in order to find suitable geometric explanation of this phenomenon.
 So, this collection of examples may be helpful for investigations of the
 invariant geometric properties of metric $\mathrm g$, geodesic $\Pi$ and potential $\Lambda$ matrices as objects on the whole phase space, which allows us to  obviate a necessity of the direct solutions of the equations (\ref{m-eq1},\ref{w-eq}) and (\ref{bi-inv}).
Moreover, it can possibly be a suitable step towards the construction of Poisson bivectors
on more generic symplectic and Poisson manifolds.


\begin{thebibliography}{10}



\bibitem{ben97}
S. Benenti,
\newblock{\em Intrinsic characterization of the variable
separation in the Hamilton-Jacobi equation},
 {J. Math. Phys.}, v.38, p. 6578-6602, 1997.

\bibitem{ben05}
 S.Benenti,
\newblock{\em Special symmetric two-tensors, equivalent dynamical systems, cofactor and
bi-cofactor systems}, {Acta Applicandae Mathematicae}, v.87, p. 33-91, 2005.

\bibitem{bm03}
A. V. Bolsinov, V. S. Matveev,
\newblock{\em Geometrical interpretation of Benenti systems}, J. Geom. Phys. v.44, p.489-506, 2003.

\bibitem{bm05}
A.V. Borisov, I.S. Mamaev, \newblock{\em Rigid Body Dynamics. Hamiltonian Methods, Integrability, Chaos}, Moscow-Izhevsk, RCD, 2005.

\bibitem{br}
R. Brouzet,
\newblock{\em About the existence of recursion operators for completely integrable
Hamiltonian systems near a Liouville torus}, Jour. Math. Phys. 34, 1309–1313, 1993.

\bibitem{chap03} 
S.A. Chaplygin,
\newblock{\em A new partial solution of the problem of motion of a
rigid body in a liquid}, Trudy otdel. Fiz. Nauk Obsh. Liub. Est., v.11, p.7-10, 1903.

\bibitem{chap04}
S.A. Chaplygin,
\newblock{\em A new partial solution of the problem of rotation of a heavy rigid body about a
fixed point}, Trudy otd. fiz. nauk Mosk. obshch. lyub. estest., v. 12, no. 1, p. 1–4, 1904.

\bibitem{sar00}
M. Crampin, W. Sarlet, G. Thompson,
\newblock{\em Bi-differential calculi, bi-Hamiltonian systems and conformal
Killing tensors},
J. Phys. A: Math. Gen. v.33, p.8755–8770., 2000.


\bibitem{dull04}
H.R. Dullin, V.S. Matveev,
\newblock{\em A new integrable system on the sphere},
Mathematical Research Letters, v.11, p.715-722, 2004.

\bibitem{gaff}
B. Gaffet,
\newblock{\em A completely integrable Hamiltonian motion on the surface of a sphere},
J. Phys. A: Math. Gen., v.31, p. 1581-1596, 1998.

\bibitem{gor00} 
D.N. Goryachev,
\newblock{\em On a Motion of a Heavy Rigid Body About a Fixed Point in the Case of A = B =4C},
Mat. sbonik kruzhka lyub. mat. nauk, vol. 21, no. 3, pp. 431-438, 1900.

\bibitem{gor15} 
D.N. Goryachev,
\newblock{\em New cases of a rigid body motion about a fixed point,}
Warshav. Univ. Izv., v.3, p.1-11, 1915.

\bibitem{gor16} 
D.N. Goryachev,
\newblock{\em New cases of integrability of Euler's dynamical equations}, Warshav. Univ. Izv.,
v.3, p.1-15, 1916.

\bibitem{gts10}
Yu. A. Grigoryev, A. V. Tsiganov,
\newblock{\em Separation of variables for the generalized Henon-Heiles system and system with quartic potential},
arXiv:1012.0468, 2010.

\bibitem{imm00}
A. Ibort, F. Magri, G. Marmo,
\newblock {\em Bihamiltonian structures and St\"{a}ckel separability},
 {J. Geometry and Physics}, v.33, p.210-228, 2000.

 \bibitem{kst03} I.V. Komarov, V.V. Sokolov, A.V. Tsiganov,
 \newblock{\em Poisson maps and integrable deformations of Kowalevski top.},
 \newblock{ J. Phys. A.}, v.36, p. 8035-8048, 2003.

\bibitem{kow89}
S. Kowalevski,
\newblock{\em Sur le probl\'{e}me de la rotation d'un corps solide
autour d'un point fixe},
\newblock{\em Acta Math.}, v.{12}, p.177-232, 1889.

 \bibitem{mpt}
A. J.~Maciejewski, M.~ Przybylska, A.V.~ Tsiganov,
\newblock{\em On a certain algebraic construction of integrable and superintegrable
systems}, 	arXiv:1011.3249, 2010.

\bibitem{mat10}
V. S. Matveev, V.V. Shevchishin,
\newblock{\em Differential invariants for cubic integrals of geodesic flows on surfaces},
Journal of Geometry and Physics, v. 60, pp. 833-856 , 2010.\\
\newblock{\em Two-dimensional superintegrable metrics with one linear and one cubic integral},
	arXiv:1010.4699, 2010.

 \bibitem{sok}
 V.V. Sokolov,
 \newblock{\em A new integrable case for the Kirchhoff equation},
 Theor. Math. Phys., v.129, p. 1335-1340, 2001.


 \bibitem{ts99b} A.V. Tsiganov,
 \newblock{\em On an integrable deformation of the spherical top},
J.Phys.A., v.32, p.8355-8363, 1999


 \bibitem{ts02} A.V. Tsiganov,
 \newblock{\em On the Kowalevski-Goryachev-Chaplygin gyrostat},
J. Phys. A, Math. Gen. 35, No.26, L309-L318, 2002

 \bibitem{ts05} A.V. Tsiganov,
\newblock{\em On a family of integrable systems on $S^2$ with a cubic integral of motion},
J. Phys. A, Math. Gen. v.38, p.921-927, 2005.

\bibitem{ts05a} A.V. Tsiganov,
\newblock{\em 	On integrable system on the sphere with the second integral quartic in the momenta},
J. Phys. A, Math. Gen. v.38, p.3547-3553, 2005.

\bibitem{ts06}
A.V. Tsiganov,
\newblock{\em	A note on elliptic coordinates on the Lie algebra e(3)},
J. Phys. A, Math. Gen. v.39, p.L571-L574, 2006.


\bibitem{ts07a}
A.V. Tsiganov,	\newblock{\em On the two different bi-Hamiltonian structures for the Toda lattice}, Journal of Physics A: Math. Theor. v.40, pp. 6395-6406, 2007.

 \bibitem{ts08} A.V. Tsiganov,	\newblock{\em On bi-Hamiltonian geometry of the Lagrange top},
J. Phys. A: Math. Theor., v.41, 315212 (12pp), 2008.

\bibitem{ts08t}
A. V. Tsiganov,
\newblock{\em The Poisson bracket compatible with the classical reflection equation algebra},
Regular and Chaotic Dynamics, v.13(3), p.191-203, 2008.

\bibitem{ts10a}
A.V. Tsiganov,
\newblock{\em On the generalized Chaplygin system},
Journal of Mathematical Sciences, v.168, n.8, p.901-911, 2010.

\bibitem{ts10k}
A.V. Tsiganov, \newblock{\em New variables of separation for particular case of the Kowalevski top},
Regular and Chaotic Dynamics, v.15, n.6, p. 657-667, 2010.

 \bibitem{ts10d} A.V. Tsiganov,
 \newblock{\em On bi-integrable natural Hamiltonian systems on the Riemannian manifolds},
 arXiv:1006.3914, 2010.



\bibitem{Turiel}
F. Turiel,
{\em Structures bihamiltoniennes sur le fibr\'e cotangent},
C.\ R.\ Acad.\ Sci.\ Paris S\'er.\ I Math.\ {\bf 315} (1992),
1085--1088.

\bibitem{val10}
G. Valent, \newblock{\em On a class of integrable systems with a cubic first integral},
Commun. Math. Phys., v.299, p.631-649, 2010.

\bibitem{ts09v}
A.V. Vershilov, A.V. Tsiganov,
\newblock{\em
On bi-Hamiltonian geometry of some integrable systems on the sphere with cubic integral of motion}, J. Phys. A: Math. Theor. v.42, 105203 (12pp), 2009.

\bibitem{yeh}
H.M. Yehia, A.A. Elmandouh,\newblock{\em New integrable systems with a quartic integral and new generalizations of Kovalevskaya's and Goriatchev's cases},
Regular and Chaotic Dynamics, v.13(1), pp. 56 - 69, 2008.

\end{thebibliography}
\end{document}